\documentclass[preprint,pre,aps,amsfonts,amsmath,showpacs,draft]{revtex4}
\begin{document}
\title{Note on a diffraction-amplification problem}
\author{Philippe Mounaix}
\email{mounaix@cpht.polytechnique.fr}
\affiliation{Centre de Physique Th\'eorique, UMR 7644 du CNRS, Ecole
Polytechnique, 91128 Palaiseau Cedex, France.}
\author{Joel L. Lebowitz}
\email{lebowitz@math.rutgers.edu}
\affiliation{Departments of Mathematics and Physics, Rutgers, The State
University of New Jersey, Piscataway, New Jersey 08854-8019.}
\date{\today}
\begin{abstract}
We investigate the solution of the equation
$\partial_t{\cal E}(x,t)-i{\cal D}\partial_x^2 {\cal E}(x,t)=
\lambda\vert S(x,t)\vert^2{\cal E}(x,t)$, for $x$ in a circle and
$S(x,t)$ a Gaussian stochastic field with a
covariance of a particular form. It is shown that
the coupling $\lambda_c$ at which $\langle\vert {\cal E}\vert\rangle$
diverges for $t\ge 1$ (in suitable units), is always less or equal for
${\cal D}>0$ than ${\cal D}=0$.
\end{abstract}
\maketitle
\section{Introduction}
\label{sec1}
In a recent work, Asselah, DaiPra, Lebowitz, and Mounaix (ADLM)\
\cite{ADLM} analyzed the divergence of the average solution to the
following diffusion-amplification problem
\begin{equation}\label{eq0.1}
\left\lbrace\begin{array}{l}
\partial_t{\cal E}(x,t)-{\cal D}\Delta
{\cal E}(x,t)=
\lambda S(x,t)^2{\cal E}(x,t),\\
t\geq 0,\ x\in\Lambda\subset {\mathbb R}^d,\ {\rm and}\
{\cal E}(x,0)=1.
\end{array}\right.
\end{equation}
Here ${\cal D} \geq 0$ is the diffusion constant, $\Lambda$ is a
$d$-dimensional torus, $\lambda >0$ is a coupling constant to the
statistically homogeneous Gaussian driver field $S(x,t)$ with $\langle
S(x,t)\rangle =0$ and $\langle S(x,t)^2\rangle =1$. They proved that,
under some reasonable assumptions on the covariance of $S$, the average
solution of $(1)$ with $D > 0$ diverges at an earlier (or equal) time
than when $D=0$.  Put otherwise, fix $T$ such that $\langle {\cal
E}(x,T) \rangle = \infty$ for $\lambda >\lambda_c$ and $\langle {\cal
E}(x,T) \rangle < \infty$ for $\lambda <\lambda_c$. Then $\lambda_c$ is
smaller than (or equal to) $\overline{\lambda}_c$ the value of
$\lambda$ at which such a divergence occurs for ${\cal D}=0$. ADLM
conjectured that this result should also apply to the case  where
${\cal D}$ is replaced by $i{\cal D}$, i.e.\  where diffusion is
replaced by diffraction, the case of physical interest considered by
Rose and DuBois in Ref.\ \cite{RD}.
\paragraph*{}The difficulty in proving the above conjecture lies in
controlling the complex Feynman path-integral, compared to that of the
Feynman-Kac formula for the diffusive case. One cannot {\it a priori}
exclude the possibility that destructive interference effects between
different paths make the sum of divergent contributions finite, raising
the value of the coupling constant at which the average amplification
diverges.To understand this diffraction-induced interference between
paths, we investigate here the diffraction case in a one dimensional model
($d=1$) in which the Gaussian driver field $S$ has a special form specified
in Section\ \ref{sec2}. We prove in Section\ \ref{sec3} that $\langle\vert {\cal E}(x,T)
\vert\rangle = \infty$ for $\lambda >\lambda_c$ with
$\lambda_c\le\overline{\lambda}_c$. Possible generalizations are discussed
in Section\ \ref{sec4}.
%
%
\section{Model and definitions}
\label{sec2}
We consider the diffraction-amplification equation
\begin{equation}\label{eq1.1}
\left\lbrace\begin{array}{l}
\partial_t{\cal E}(x,t)-\frac{i}{2}\Delta
{\cal E}(x,t)=
\lambda\vert S(x,t)\vert^2{\cal E}(x,t),\\
x\in\Lambda_1, \ {\rm and}\
{\cal E}(x,0)=1,
\end{array}\right.
\end{equation}
where $\lambda >0$ is the coupling constant and $\Lambda_1$ is a circle
of unit circumference. The case in which the circle has circumference
$L$ and/or there is a constant ${\cal D}$ multiplying $\Delta {\cal E}$
is straightforwardly obtained by rescaling $x$, $t$, and $\lambda$. The
driver amplitude $S(x,t)$ is a space time homogeneous complex Gaussian
random field with
\begin{equation}\label{jll1}
\left\lbrace\begin{array}{l}
\langle S(x,t)\rangle =\langle S(x,t)S(x',t')\rangle =0, \\
\langle S(x,t)S^\ast (x',t')\rangle =C(x-x',t-t'),
\end{array}\right.
\end{equation}
and $C(0,0)=1$. We can write $S(x,t)$ in the form
\begin{equation}\label{eq1.2}
S(x,t)=\sum_{n\in {\mathbb Z}}
\xi_n(t) {\rm e}^{2i\pi nx},
\end{equation}
with $\xi_n(t)$ Gaussian random functions satisfying
\begin{equation}\label{jll2}
\left\lbrace\begin{array}{l}
\langle\xi_n(t)\rangle =\langle \xi_n(t)\xi_m(t')\rangle =0, \\
\langle \xi_n(t)\xi_m^\ast (t')\rangle =\delta_{nm}C_n(t-t'),
\end{array}\right.
\end{equation}
with $C_n(0)\equiv\epsilon_n\ge 0$ and $\sum\epsilon_n =1$. We now assume
that only a finite number of $\epsilon_n$ are non vanishing;
\begin{equation}\label{eq1.4}
\epsilon_n = 0 \quad {\rm for\ } |n| > N, \quad N < \infty ,
\end{equation}
reducing the right-hand side (rhs) of Eq.\ (\ref{eq1.2}) to a finite sum of
$M=2N+1$ terms, from $n=-N$ to $n=N$. We further assume that
\begin{equation}\label{jll3}
\xi_n(t)=\sqrt{\epsilon_n}\phi_n(t)s_n,
\end{equation}
where the $\phi_n(t)$ are specified functions of $t$ and the $s_n$ are
independent complex Gaussian random variables with
\begin{equation}\label{eq1.3}
\left\lbrace\begin{array}{l}
\langle s_n\rangle =\langle s_n s_m\rangle =0,\\
\langle s_n s_m^\ast\rangle =\delta_{nm}.
\end{array}\right.
\end{equation}
It then follows from\ (\ref{jll2}),\ (\ref{jll3}), and\ (\ref{eq1.3}) that
\begin{equation}\label{jll4}
\phi_n(t)=\exp(i\omega_n t), \quad \omega_n\ {\rm real},
\end{equation}
yielding
\begin{equation}\label{jll5}
C(x-x',t-t')=\sum_{n=-N}^N \epsilon_n
{\rm e}^{i\lbrack 2\pi n(x-x') +\omega_n(t-t')\rbrack}.
\end{equation}
In the following we take $\omega_n =an^2$, $a>0$, which is the case of interest in
optics where the space-time behavior of $C(x,t)$ corresponds to a diffraction
along $x$ as $t$ increases. The last and most restrictive assumption we make is
that the $\phi_{n\ge 0}(t)$ are orthogonal functions of $t$ in $\lbrack 0,1\rbrack$,
which specifies $a$. One finds
\begin{equation}\label{jll6}
\omega_n=2\pi n^2,\quad {\rm i.e.\ }\phi_n(t)=\exp(2i\pi n^2 t).
\end{equation}
Equation\ (\ref{eq1.1}) can thus be rewritten as
\begin{equation}\label{eq1.1bis}
\partial_t{\cal E}(x,t)-\frac{i}{2}\Delta
{\cal E}(x,t)=
\lambda\, s^{\dag}\gamma (x,t)s\, {\cal E}(x,t),
\end{equation}
where $s$ is the $M$-line Gaussian random vector the elements of which
are the $s_n$, and $\gamma (x,t)$ is a $M\times M$
Hermitian matrix with elements
\begin{equation}\label{eq1.gam}
\gamma_{nm}(x,t)=\sqrt{\epsilon_n\epsilon_m}{\rm e}^{-2i\pi\lbrack
(n-m)x+(n^2-m^2)t\rbrack}.
\end{equation}
\paragraph*{}Finally, the critical coupling $\lambda_c$ and its
diffraction-free counterpart $\overline{\lambda}_c$ are defined by
\begin{subequations}\label{eq1.5}
\begin{eqnarray}
&&\lambda_c=
\inf\lbrace \lambda>0:
\langle\vert {\cal E}(0,1)\vert\rangle =+\infty\rbrace,
\label{eq1.5a}\\
&&\overline{\lambda}_c=
\inf\lbrace \lambda>0:\langle {\rm e}^{\lambda\int_0^1
S(0,t)^2dt}\rangle =+\infty\rbrace,
\label{eq1.5b}
\end{eqnarray}
\end{subequations}
where $\langle .\rangle$ denotes the average over the realizations of
$S$. Equations\ (\ref{eq1.5}) give the values of $\lambda$ at which
$\langle\vert {\cal E}(x,t)\vert\rangle$ diverges after one unit of time
with and without diffraction respectively.
%
%
\section{Comparison of $\lambda_c$ and $\overline{\lambda}_c$}
\label{sec3}
We begin with two  lemmas that will be useful in the following.
Let ${\cal E}_{\gamma}(x,t)$ be the solution to Eq.\ (\ref{eq1.1bis})
for a given realization of $s$.
\paragraph*{}
\bigskip
\noindent {\bf Lemma 1.}
For every $x\in{\mathbb R}$ and $t\in\lbrack
0,1\rbrack$, and every $M\times M$ unitary matrix $P$, one has
$\langle\vert {\cal E}_{\gamma}(x,t)\vert\rangle = \langle\vert {\cal
E}_{P^{\dag}\gamma P}(x,t)\vert\rangle$.
\paragraph*{}
\bigskip
\noindent {\bf Proof.}
Let $B(x,t)$ be the set of all the continuous paths $x(\tau)$, with
$t\in\lbrack 0,1\rbrack$, $\tau\le t$, and $x(\tau)\in {\mathbb R}$,
arriving at $x(t)=x$. Writing the solution to Eq.\ (\ref{eq1.1bis}) as
a Feynman path-integral, one has
\begin{eqnarray*}
&&\langle\vert {\cal E}_{\gamma}(x,t)\vert\rangle =
\int_{{\mathbb C}^M}\frac{{\rm e}^{-\vert s\vert^2}}{\pi^M}
\left\vert
\int_{x(\cdot)\in B(x,t)}{\rm e}^{\int_0^t\left\lbrack\frac{i}{2}
\dot{x}(\tau)^2 +\lambda s^{\dag}\gamma(x(\tau),\tau)s\right\rbrack
d\tau}d\lbrack x(\cdot)\rbrack\right\vert\prod_nd^2s_n \\
&&=\int_{{\mathbb C}^M}\frac{{\rm e}^{-s^{\dag}PP^{\dag}s}}{\pi^M}
\left\vert
\int_{x(\cdot)\in B(x,t)}{\rm e}^{\int_0^t\left\lbrack\frac{i}{2}
\dot{x}(\tau)^2 +\lambda
s^{\dag}PP^{\dag}\gamma(x(\tau),\tau)PP^{\dag}s\right\rbrack
d\tau}d\lbrack x(\cdot)\rbrack\right\vert\prod_nd^2s_n \\
&&=\int_{{\mathbb C}^M}\frac{{\rm e}^{-\vert\sigma\vert^2}}{\pi^M}
\left\vert
\int_{x(\cdot)\in B(x,t)}{\rm e}^{\int_0^t\left\lbrack\frac{i}{2}
\dot{x}(\tau)^2 +\lambda
\sigma^{\dag}P^{\dag}\gamma(x(\tau),\tau)P\sigma\right\rbrack
d\tau}d\lbrack x(\cdot)\rbrack\right\vert\prod_nd^2\sigma_n
=\langle\vert {\cal E}_{P^{\dag}\gamma P}(x,t)\vert\rangle .
\end{eqnarray*}
Here we have used $PP^{\dag}=1$ and made the change of variables
$s_n\rightarrow\sigma_n$ where the $\sigma_n$ are the components of
$\sigma\equiv P^{\dag}s$. Note that Lemma 1 applies also to the
diffraction-free case by eliminating the path integral and
setting $x(\tau)\equiv x$.
\paragraph*{}
\bigskip
Let $\kappa_n$ ($n\in {\mathbb N}$) be the eigenvalues of the $M\times
M$ Hermitian matrix $\int_0^1\gamma(0,t)\, dt$. One has the following
Lemma
\paragraph*{}
\bigskip
\noindent {\bf Lemma 2.}
$\overline{\lambda}_c=(\sup_n\kappa_n)^{-1}$.
\paragraph*{}
\bigskip
\noindent {\bf Proof.}
Using Eq.\ (\ref{eq1.gam}) one finds, after a suitable permutation of
lines and columns, that $\int_0^1\gamma(0,t)\, dt$ can be written in
the block-diagonal form
\begin{equation}\label{eq2.1}
\int_0^1\gamma(0,t)\, dt=\left(
\begin{array}{ccccc}
\epsilon_0&0&\cdots& &\\
0&g_1&0&\cdots&\\
\vdots&0&\ddots&0&\cdots\\
 &\vdots&0&g_{N-1}&0\\
 & &\vdots&0&g_N
\end{array}\right),
\end{equation}
with
\begin{equation}\label{eq2.1bis}
g_j=\left(
\begin{array}{cc}
\epsilon_j&\sqrt{\epsilon_j\epsilon_{-j}}\\
\sqrt{\epsilon_j\epsilon_{-j}}&\epsilon_{-j}
\end{array}\right),
\end{equation}
the diagonalization of which yields the $M$ eigenvalues $\kappa_n$.
These eigenvalues are easily found to be $\epsilon_0$,
$\epsilon_j+\epsilon_{-j}$, and $0$. The matrix diagonalizing\
(\ref{eq2.1}), $P$, is a unitary matrix given by
\begin{equation}\label{eq2.2}
P=\left(
\begin{array}{ccccc}
1&0&\cdots& &\\
0&p_1&0&\cdots&\\
\vdots&0&\ddots&0&\cdots\\
 &\vdots&0&p_{N-1}&0\\
 & &\vdots&0&p_N
\end{array}\right),
\end{equation}
with
\begin{equation}\label{eq2.2bis}
p_j=\left(
\begin{array}{cc}
\sqrt{\epsilon_j/(\epsilon_j+\epsilon_{-j})}&
\sqrt{\epsilon_{-j}/(\epsilon_j+\epsilon_{-j})}\\
\sqrt{\epsilon_{-j}/(\epsilon_j+\epsilon_{-j})}&
 -\sqrt{\epsilon_j/(\epsilon_j+\epsilon_{-j})}
\end{array}\right).
\end{equation}
Using the diffraction-free version of Lemma 1 with $P$ given by Eqs.\
(\ref{eq2.2}) and\ (\ref{eq2.2bis}), one obtains
\begin{eqnarray}\label{eq2.3}
\langle {\rm e}^{\lambda\int_0^1 S(0,t)^2dt}\rangle &=&
\int_{{\mathbb C}^M}\frac{{\rm e}^{-\vert\sigma\vert^2}}{\pi^M}
{\rm e}^{\lambda\sigma^{\dag}\left\lbrack\int_0^1 P^{\dag}
\gamma(0,t)P\, dt\right\rbrack\sigma}\prod_nd^2\sigma_n \nonumber \\
&=&\prod_n\int_0^{+\infty}
{\rm e}^{(\lambda\kappa_n-1)u_n} du_n ,
\end{eqnarray}
with $u_n\equiv\vert\sigma_n\vert^2$, from which Lemma 2 follows
straightforwardly. One can now prove the proposition:
\paragraph*{}
\bigskip
\noindent {\bf Proposition.} $\lambda_c\le\overline{\lambda}_c$.
\paragraph*{}
\bigskip
\noindent {\bf Proof.}
From Lemma 1 with $P$ given by Eqs.\ (\ref{eq2.2}) and\
(\ref{eq2.2bis}), one has
\begin{equation}\label{eq2.4}
\langle\vert {\cal E}(0,1)\vert\rangle =
\int_{{\mathbb C}^M}\frac{{\rm e}^{-\vert\sigma\vert^2}}{\pi^M}
\left\vert
\int_{x(\cdot)\in B(0,1)}{\rm e}^{\int_0^1\left\lbrack\frac{i}{2}
\dot{x}(\tau)^2 +\lambda
\sigma^{\dag}P^{\dag}\gamma(x(\tau),\tau)P\sigma\right\rbrack
d\tau}d\lbrack x(\cdot)\rbrack\right\vert\prod_nd^2\sigma_n .
\end{equation}
For this integral to exist  it is necessary that
\begin{equation}\label{eq2.5}
\lim_{\vert\sigma\vert\rightarrow +\infty}
{\rm e}^{-\vert\sigma\vert^2}
\left\vert
\int_{x(\cdot)\in B(0,1)}{\rm e}^{\int_0^1\left\lbrack\frac{i}{2}
\dot{x}(\tau)^2 +\lambda
\sigma^{\dag}P^{\dag}\gamma(x(\tau),\tau)P\sigma\right\rbrack
d\tau}d\lbrack x(\cdot)\rbrack\right\vert =0,
\end{equation}
for all the directions $\sigma /\vert\sigma\vert$ in ${\mathbb C}^M$.
We will now show that this cannot happen for $\lambda \ge \bar \lambda_c$.
Let $ \kappa_m =\sup_n\kappa_n$. From Lemma 2 one has
$\kappa_m =1/\overline{\lambda}_c$. Now, consider Eq.\ (\ref{eq2.5})
for $\sigma_n = 0, n \ne m$ and  $\sigma_m=z\in
{\mathbb C}$. One finds after some straightforward algebra
\begin{equation}\label{eq2.7a}
\sigma^{\dag}P^{\dag}\gamma(x,t)P\sigma =
\left\lbrack\frac{1}{\overline{\lambda}_c}
-\alpha_m\overline{\lambda}_c\sin^2(2\pi kx)
\right\rbrack\vert z\vert^2,
\end{equation}
and
\begin{eqnarray}
&&{\rm e}^{-\vert\sigma\vert^2}
\left\vert
\int_{x(\cdot)\in B(0,1)}{\rm e}^{\int_0^1\left\lbrack\frac{i}{2}
\dot{x}(\tau)^2 +\lambda
\sigma^{\dag}P^{\dag}\gamma(x(\tau),\tau)P\sigma\right\rbrack
d\tau}d\lbrack x(\cdot)\rbrack\right\vert \nonumber \\
&&={\rm e}^{(\lambda/\overline{\lambda}_c -1)\vert z\vert^2}
\left\vert
\int_{x(\cdot)\in B(0,1)}{\rm e}^{\int_0^1\left\lbrack\frac{i}{2}
\dot{x}(\tau)^2 -\lambda\vert z\vert^2\alpha_m\overline{\lambda}_c
\sin^2\left(2\pi kx(\tau)\right)\right\rbrack
d\tau}d\lbrack x(\cdot)\rbrack\right\vert ,\label{eq2.7}
\end{eqnarray}
where $\alpha_m = 4\epsilon_k\epsilon_{-k}$ if $\kappa_m =
\epsilon_k+\epsilon_{-k}$, which defines $k$, and $\alpha_m = 0$ if
$\kappa_m = \epsilon_0$. There are two possibilities:
\paragraph*{}(i) If $\alpha_m = 0$ one has
\begin{eqnarray}
&&{\rm e}^{-\vert\sigma\vert^2}
\left\vert
\int_{x(\cdot)\in B(0,1)}{\rm e}^{\int_0^1\left\lbrack\frac{i}{2}
\dot{x}(\tau)^2 +\lambda
\sigma^{\dag}P^{\dag}\gamma(x(\tau),\tau)P\sigma\right\rbrack
d\tau}d\lbrack x(\cdot)\rbrack\right\vert \nonumber \\
&&={\rm e}^{(\lambda/\overline{\lambda}_c -1)\vert z\vert^2}
\left\vert\int_{x(\cdot)\in B(0,1)}{\rm e}^{\int_0^1\frac{i}{2}
\dot{x}(\tau)^2 d\tau}d\lbrack x(\cdot)\rbrack\right\vert
={\rm e}^{(\lambda/\overline{\lambda}_c -1)\vert z\vert^2}.\label{eq2.6}
\end{eqnarray}
If $\lambda_c >\overline{\lambda}_c$ this expression diverges as $\vert
z\vert$ tends to infinity, which is in contradiction with Eq.\
(\ref{eq2.5}).
\paragraph*{}(ii) If $\alpha_m\ne 0$ the leading term of the asymptotic
expansion of the path-integral\ (\ref{eq2.7}) in the large $\vert
z\vert$ limit is given by the contribution of the paths near
$x(\tau)=0$. Expanding $\sin^2(2\pi kx)$ around $x=0$ at the lowest
order and performing the resulting Gaussian integral, one obtains the
asymptotics
\begin{eqnarray}
&&{\rm e}^{-\vert\sigma\vert^2}
\left\vert
\int_{x(\cdot)\in B(0,1)}{\rm e}^{\int_0^1\left\lbrack\frac{i}{2}
\dot{x}(\tau)^2 +\lambda
\sigma^{\dag}P^{\dag}\gamma(x(\tau),\tau)P\sigma\right\rbrack
d\tau}d\lbrack x(\cdot)\rbrack\right\vert \nonumber \\
&&\sim\sqrt{2}{\rm e}^{(\lambda/\overline{\lambda}_c -1)\vert z\vert^2}
{\rm e}^{-\vert z\vert\pi k\sqrt{\alpha_m\lambda\overline{\lambda}_c}}
\ \ \ \ (\vert z\vert\rightarrow +\infty).\label{eq2.8}
\end{eqnarray}
Again, if $\lambda_c >\overline{\lambda}_c$ the rhs of
this expression diverges as $\vert z\vert$ tends to infinity, which
completes the proof of the proposition.
%
%
\section{Discussion and perspectives}\label{sec4}
As a conclusion we would like to outline a possible way of fitting the
ideas behind this calculation to a more general proof of the
conjecture. First, it should be noticed that what makes the proof here
possible is the slow decrease of the asymptotic behavior of the path
integral on the rhs of Eq.\ (\ref{eq2.7}) as $\vert z\vert\rightarrow
+\infty$. Namely, denoting by $f(\vert z\vert)$ this path integral,
one has $\forall\varepsilon > 0$, $\lim_{\vert z\vert\rightarrow
+\infty} \vert f(\vert z\vert)\vert\exp(\varepsilon\vert
z\vert^2)=+\infty$ [cf. Eqs.\ (\ref{eq2.6}) and\ (\ref{eq2.8})], which
proves the conjecture  by leading to a contradiction with
Eq.\ (\ref{eq2.5}).
\paragraph*{}Now, consider the case in which $S(x,t)$ is given by a
finite Karhunen-Lo\`eve type expansion $S(x,t)=\sum_n s_n\Phi_n(x,t)$
with $x\in {\mathbb R}^d$, $t\in\lbrack 0,T\rbrack$, and $\Phi_n(x,t)$
not necessarily periodic in time\ \cite{note1}. With such an expression
for $S(x,t)$ on the rhs of Eq.\ (\ref{eq0.1}), one finds that the
equation for ${\cal E}(x,t)$ takes on the same form as in\
(\ref{eq1.1bis}) with $\gamma_{nm}(x,t)=\Phi_n(x,t)\Phi_m(x,t)^\ast$.
One could now 
systematically replace, from Eq.\ (\ref{eq2.4}) on, the matrix
diagonalizing $\int_0^1\gamma(0,t)\, dt$ by the one diagonalizing
$\Gamma\lbrack y(\cdot)\rbrack\equiv\int_0^T\gamma(y(t),t)\, dt$, where
$y(\cdot)\in B(0,T)$ is a continuous path maximizing the largest
eigenvalue of $\Gamma\lbrack x(\cdot)\rbrack$\ \cite{note2}. Denoting
by $\kappa_c$ this maximized largest eigenvalue, one expects the rhs of
Eq.\ (\ref{eq2.7}) to be replaced by
\begin{equation}\label{eq3.1}
{\rm e}^{(\lambda\kappa_c -1)\vert z\vert^2}
\left\vert
\int_{x(\cdot)\in B(0,T)}{\rm e}^{\int_0^T\left\lbrack\frac{i}{2}
\dot{x}(\tau)^2 -\lambda\vert z\vert^2V(x(\tau),\tau)\right\rbrack
d\tau}d\lbrack x(\cdot)\rbrack\right\vert ,
\end{equation}
where $V(x,t)$ is a real potential given by some
linear combination of the $\gamma_{nm}(x,t)$ and such that
\begin{equation}\label{eq3.2}
\inf_{x(\cdot)\in B(0,T)}\int_0^T V(x(\tau),\tau)\, d\tau =0.
\end{equation}
The proof would then
proceed along exactly the same line as in this note: denote by $f(\vert
z\vert)$ the path integral in Eq.\ (\ref{eq3.1}), if one can prove that
$\forall\varepsilon > 0$, $\lim_{\vert z\vert\rightarrow +\infty} \vert
f(\vert z\vert)\vert\exp(\varepsilon\vert z\vert^2)>0$ (which seems to
be the difficult part of the matter), then we will have proved
$\lambda_c\le\kappa_c^{-1}$. Finally, since $1/\overline{\lambda}_c$ is
the largest eigenvalue of $\Gamma\lbrack x(\cdot)=0\rbrack$, it is
necessarily smaller than (or equal to) $\kappa_c$, and
$\lambda_c\le\kappa_c^{-1}$ implies $\lambda_c\le\overline{\lambda}_c$.
Note that Eq.\ (\ref{eq2.7}) is a particular case of Eqs.\ (\ref{eq3.1})
and\ (\ref{eq3.2}) with $\kappa_c=1/\overline{\lambda}_c$ and $V(x,t)=
\alpha_m\overline{\lambda}_c\sin^2(2\pi kx)$. 
\begin{acknowledgments}
We thank Pierre Collet for many useful discussions.  The work of JLL was
supported by AFOSR Grant AF 49620-01-1-0154, and NSF Grant DMR
01-279-26.
\end{acknowledgments}
%
%

%

\begin{thebibliography}{1}
\bibitem{note1} Note that\ (\ref{eq1.2}) belongs to this class of driver
with $d=1$ and $\Phi_n(x,t)=\sqrt{\epsilon_n}\exp\lbrack 2i\pi
(nx+n^2t)\rbrack$.
\bibitem{note2} In the cases where there is no such a path, one should
consider a path which realizes the supremum up to a arbitrarily small
constant.
\end{thebibliography}
\end{document}